\documentclass[conference]{IEEEtran}
\usepackage{cite}
\usepackage{amsmath,amssymb,amsfonts}
\usepackage{algorithmic}
\usepackage{graphicx}
\usepackage{textcomp}
\usepackage{xcolor}
\def\BibTeX{{\rm B\kern-.05em{\sc i\kern-.025em b}\kern-.08em
    T\kern-.1667em\lower.7ex\hbox{E}\kern-.125emX}}
\begin{document}

\title{Scaling Performance of Large Language Model Pretraining}

\author{
\IEEEauthorblockN{
    Alexander Interrante-Grant,
    Carla Varela-Rosa,
    Suhaas Narayan,
    Chris Connelly,
    Albert Reuther
}
\IEEEauthorblockA{
    \textit{Lincoln Laboratory} \\
    \textit{Massachusetts Institute of Technology} \\
    Lexington, MA, USA \\
    \{ainterr, carla.varela-rosa, suhaas.narayan, connelly, reuther\}@ll.mit.edu
}
}

\maketitle

\begin{abstract}
Large language models (LLMs) show best-in-class performance across a wide range
of natural language processing applications. Training these models is an
extremely computationally expensive task; frontier Artificial Intelligence (AI)
research companies are investing billions of dollars into supercomputing
infrastructure to train progressively larger models on increasingly massive
datasets. Unfortunately, very little information about the scaling performance
and training considerations of these large training pipelines is released
publicly. Working with very large datasets and models can be complex and
practical recommendations are scarce in the public literature for tuning
training performance when scaling up large language models. In this paper, we
aim to demystify the large language model pretraining pipeline somewhat - in
particular with respect to distributed training, managing large datasets across
hundreds of nodes, and scaling up data parallelism with an emphasis on fully
leveraging available GPU compute capacity.
\end{abstract}


\section{Introduction}

Scaling up LLMs has been the primary goal of most frontier AI research
companies over the past several years \cite{openai-2025-gpt45,
google-2025-gemini25, anthropic-2024-claude35, xai-2025-grok3,
mistral-2025-medium3, deepseek-2025-r1}. Recent research on LLMs hash shown
that, given sufficient data, simply scaling up the model size can predictably
improve performance across a wide range of downstream tasks
\cite{wei-2022-emergence}. However, larger models require more computational
resources to train. Further, larger datasets also increase the computational
burden of training these models. It's no surprise then that companies like
Google, OpenAI (via Microsoft), Anthropic, and others have invested billions of
dollars into supercomputing infrastructure to produce some of the world's most
advanced AI models.

\begin{table}[htbp]
\caption{Frontier Models}
\centering
\begin{center}
\begin{tabular}{|c|c|c|}
\hline
\textbf{Company}&\textbf{Model}&\textbf{Release Date} \\
\hline
OpenAI & GPT-4.5 \cite{openai-2025-gpt45} & February, 2025 \\
Google & Gemini 2.5 \cite{google-2025-gemini25} & July, 2025 \\
Anthropic & Claude 3.5 Sonnet \cite{anthropic-2024-claude35} & June, 2024 \\
xAI & Grok 3 \cite{xai-2025-grok3} & February, 2025 \\
Mistral AI & Medium 3 \cite{mistral-2025-medium3} & May, 2025 \\
DeepSeek & R1 \cite{deepseek-2025-r1} & January, 2025 \\
\hline
\end{tabular}
\label{table:frontier-models}
\end{center}
\end{table}

Recent model releases from frontier AI companies (see Table
\ref{table:frontier-models}) reveal very little about their model architectures
or training processes. Even basic information such as model sizes (parameter
counts) are rarely reported publicly and left to rumor and speculation. This
leaves researchers training similar models to resolve issues with scaling
(distributed data access, bottleneck resolution, best practices) on their own.

In this work, we pretrain an LLM for a novel language application, examining
scaling performance as we increase the size of both the dataset and the model,
and report lessons learned along the way. All of this work was done on a
computing cluster of several hundred nodes equipped with Nvidia H100 GPUs.

\subsection{Supercomputing Infrastructure}

The Lincoln Laboratory Supercomputing Center (LLSC) operates, maintains, and
supports thousands of users operating at a variety of security levels. The
Nvidia H100 nodes on which this work was completed is the TX-GAIN (GenAI Next)
system, one of several partitions/clusters that comprise the LLSC's TX-Green
supercomputer. TX-GAIN is a cluster of 316 HPE compute nodes with dual AMD EPYC
9254 24-core CPUs with 768 GB of DRAM and dual Nvidia Hopper H100-NVL GPUs with
94 GB of HBM GPU RAM. The H100-NVL GPUs have a NV-Link bridge that interconnect
the two GPUs on each compute node. Each node mounts the central Lustre parallel
storage array file systems, has 3.8 TB of local SSD storage, and includes a
25-Gigabit Converged Ethernet link to a non-blocking central Converged Ethernet
cluster core switch. The TX-GAIN cluster was deployed in early summer 2025, and
it is \#114 on the June 2025 Top500 List.

\section{Scaling Up Model Training}

While the model and dataset are not the subject of this paper, some details of
the model are relevant to its scaling performance. Our experiments involved
pretraining a Bidirectional Encoder Representation for Transformers (BERT)-like
encoder model \cite{devlin-2018-bert} on a large, novel dataset of binary code.
The model was pretrained on a Masked Language Modeling (MLM) task with 15\% of
tokens in the training dataset randomly masked.  As we iterated on the model,
we scaled from experimental training runs of a small 120M parameter model on a
small shard of the dataset running on a single GPU, all the way up to a larger
350M parameter model trained on the full dataset distributed across 128 nodes
with 256 GPUs in total.

\subsection{Dataset}

Our dataset is comprised of binary code in the form of functions (202M
pretraining samples) compiled from open-source projects using the nixpkgs
package manager \cite{nixpkgs} totaling just under 2 TB in size.  The size of
our dataset leads to our first roadblock in scaling training - sharing 2 TB of
data across hundreds of nodes in a supercomputing environment can have
considerable performance impacts. Addressing this problem leads to our first
two recommendations.

\begin{enumerate}
\item[1.] \textbf{Preprocess and tokenize the entire dataset ahead of
training}, storing only the necessary training data: tokenized inputs and
attention masks.  The bulk of the storage space for our dataset consisted of
large volumes of function data with a relatively poor compression ratio. After
tokenization and with only the fields necessary for pretraining, our dataset
was only 25 GB in size (a reduction of 99\%!).
\item[2.] If it is small enough, \textbf{duplicate your dataset across nodes
prior to training}. We found that the initial cost of copying the entire
dataset from network storage to local storage on each node was worth it to
avoid nodes contending for network and network-attached storage resources
throughout the course of training.
\end{enumerate}

These two changes allowed us to eliminate a network storage bottleneck that
would have prevented us from saturating our GPUs compute performance.

\subsection{Parallelism}

Starting with smaller training runs on a single GPU, we noticed that GPU
utilization would spike briefly and then drop to 0\% repeatedly throughout
training. This leads to our next recommendation.

\begin{enumerate}
\item[3.] \textbf{Parallelize data loading, but only just as much as
necessary}. Even though we did significant preprocessing ahead of training,
simply loading data into GPU memory was still a bottleneck. We gradually
increased the number of parallel data loaders until single GPU utilization
stabilized near 100\% - any more than this would simply be a waste of
resources\footnote{Note: we found that it was best to determine the optimal
number of data loaders \textit{after} optimizing training batch size to
saturate GPU memory.}.
\end{enumerate}

\begin{figure}[htbp]
\centerline{\includegraphics{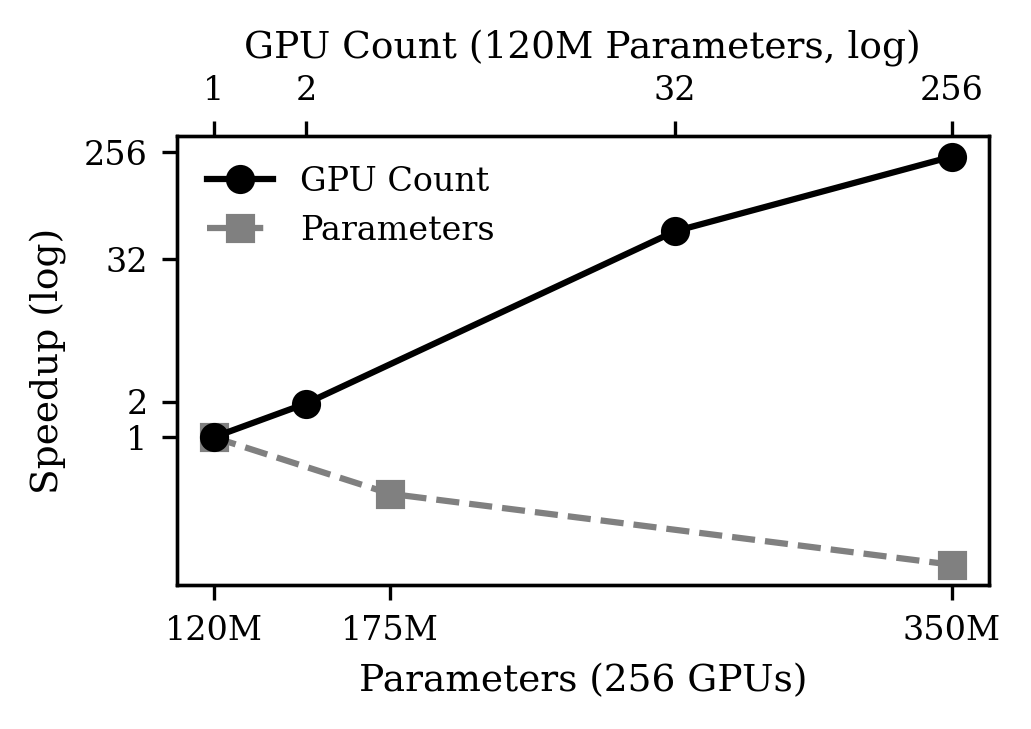}}
\caption{Pretraining scaling performance.}
\label{figure:scaling-performance}
\end{figure}

Once we were able to efficiently saturate a single GPU for training, we were
ready to scale up to multiple GPUs. We used PyTorch Lightning \cite{lightning}
to enable multi-GPU and multi-node training. Figure
\ref{figure:scaling-performance} depicts training performance across a range of
model sizes and node counts. This leads to our next insight.

\begin{enumerate}
\item[4.] For data parallel multi-node training, \textbf{network bandwidth is
not as much of a bottleneck as it might seem}. As depicted in Figure
\ref{figure:scaling-performance} - model training performance scales roughly
linearly with the number of nodes, even up to 128 data parallel nodes. This
indicates that our workload is still GPU bound and not bottlenecked by network
performance when propagating weight updates at the end of each batch.
\end{enumerate}

Lastly, as we scaled up the model size (keeping a constant 128 nodes) we
noticed a decrease in training performance.

\begin{enumerate}
\item[5.] \textbf{Larger models indirectly reduce training efficiency} with
data parallelism because the increased parameter count requires more GPU
memory, reducing training batch size. Our smallest (120M parameters) model was
trained with a batch size of 184 samples, while our largest (350M parameters)
only managed 20. Scaling any further than this would require model parallelism,
which would require further tuning to optimize fully.
\end{enumerate}

\section{Conclusion}

In this short paper we have provided some practical recommendations for
optimizing LLM pretraining performance that we learned through the process of
training our own. We hope that these recommendations are useful for other
researchers attempting to scale up custom LLM pretraining to larger models and
datasets.

\section{Distribution}

DISTRIBUTION STATEMENT A. Approved for public release: distribution unlimited.

© 2025 Massachusetts Institute of Technology

This material is based upon work supported by the Under Secretary of Defense
for Research and Engineering under Air Force Contract No. FA8702-15-D-0001. Any
opinions, findings, conclusions or recommendations expressed in this material
are those of the author(s) and do not necessarily reflect the views of the
Under Secretary of Defense for Research and Engineering. © 2025 Massachusetts
Institute of Technology. The software/firmware is provided to you on an As-Is
basis Delivered to the U.S. Government with Unlimited Rights, as defined in
DFARS Part 252.227-7013 or 7014 (Feb 2014). Notwithstanding any copyright
notice, U.S. Government rights in this work are defined by DFARS 252.227-7013
or DFARS 252.227-7014 as detailed above. Use of this work other than as
specifically authorized by the U.S. Government may violate any copyrights that
exist in this work.

\end{document}